\documentclass[sigconf]{acmart}
\usepackage{graphicx}
\usepackage{amsmath}
\usepackage{booktabs}
\usepackage{algorithm}
\usepackage{algorithmic}
\usepackage{amsfonts}
\usepackage{multirow}
\usepackage{makecell}
\usepackage{subfigure}
\usepackage{color}
\usepackage{bm}
\usepackage{epstopdf}
\usepackage{url}
\usepackage[cal=cm]{mathalfa}
\usepackage{balance}
\usepackage{threeparttable}

\makeatletter
\newcommand\notsotiny{\@setfontsize\notsotiny\@vipt\@viipt}
\makeatother

\AtBeginDocument{%
  \providecommand\BibTeX{{%
    \normalfont B\kern-0.5em{\scshape i\kern-0.25em b}\kern-0.8em\TeX}}}

\copyrightyear{2022}
\acmYear{2022}
\setcopyright{rightsretained}
\acmConference[CIKM '22]{Proceedings of the 31st ACM International Conference on Information and Knowledge Management}{October 17--21, 2022}{Atlanta, GA, USA}
\acmBooktitle{Proceedings of the 31st ACM International Conference on Information and Knowledge Management (CIKM '22), October 17--21, 2022, Atlanta, GA, USA}
\acmDOI{10.1145/3511808.3557262}
\acmISBN{978-1-4503-9236-5/22/10}

\usepackage{etoolbox}
\makeatletter
\patchcmd{\maketitle}{\@copyrightpermission}{
   \begin{minipage}{0.3\columnwidth}
     \href{http://creativecommons.org/licenses/by/4.0/}{\includegraphics[width=0.90\textwidth]{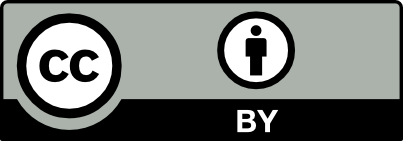}}
   \end{minipage}\hfill
   \begin{minipage}{0.7\columnwidth}
     \href{http://creativecommons.org/licenses/by/4.0/}{This work is licensed under a Creative Commons Attribution International 4.0 License.}
   \end{minipage}
  
   \vspace{5pt}
}{}{}

\author{Jiangxia Cao}
\affiliation{
  \institution{Institute of Information Engineering, Chinese Academy of Sciences  \\ School of Cyber Security, University of Chinese Academy of Sciences}
  \country{caojiangxia@iie.ac.cn}
 }

\author{Xin Cong}
\affiliation{
  \institution{Institute of Information Engineering, Chinese Academy of Sciences  \\ School of Cyber Security, University of Chinese Academy of Sciences}
  \country{congxin@iie.ac.cn}
 }

\author{Jiawei Sheng}
\affiliation{
  \institution{Institute of Information Engineering, Chinese Academy of Sciences  \\ School of Cyber Security, University of Chinese Academy of Sciences}
  \country{shengjiawei@iie.ac.cn}
 }

\author{Tingwen Liu$^\star$}
\thanks{$^\star$Tingwen Liu is the corresponding author.}
\affiliation{
  \institution{Institute of Information Engineering, Chinese Academy of Sciences  \\ School of Cyber Security, University of Chinese Academy of Sciences}
  \country{liutingwen@iie.ac.cn}
 }

\author{Bin Wang}
\affiliation{
  \institution{Xiaomi AI Lab, \\ Xiaomi Inc}
  \country{wangbin11@xiaomi.com}
}

\begin{document}
\renewcommand{\shortauthors}{Jiangxia Cao et al.}
\title{Contrastive Cross-Domain Sequential Recommendation}

\renewcommand{\shorttitle}{C$^2$DSR}

\begin{abstract}
Cross-Domain Sequential Recommendation (CDSR) aims to predict future interactions based on user's  historical sequential interactions from multiple domains.
Generally, a key challenge of CDSR is how to mine precise cross-domain user preference based on the intra-sequence and inter-sequence item interactions.
Existing works first learn single-domain user preference only with intra-sequence item interactions, and then build a transferring module to obtain cross-domain user preference.
However, such a pipeline and implicit solution can be severely limited by the bottleneck of the designed transferring module, and ignores to consider inter-sequence item relationships.
In this paper, we propose \textbf{C$^2$DSR} to tackle the above problems to capture precise user preferences.
The main idea is to simultaneously leverage the intra- and inter- sequence item relationships, and jointly learn the single- and cross- domain user preferences.
Specifically, we first utilize a graph neural network to mine inter-sequence item collaborative relationship, and then exploit sequential attentive encoder to capture intra-sequence item sequential relationship.
Based on them, we devise two different sequential training objectives to obtain user single-domain and cross-domain representations.
Furthermore, we present a novel contrastive cross-domain infomax objective to enhance the correlation between single- and cross- domain user representations by maximizing their mutual information.
To validate the effectiveness of C$^2$DSR, we first re-split four e-comerce datasets, and then conduct extensive experiments to demonstrate the effectiveness of our approach C$^2$DSR.
\end{abstract}



\maketitle

\section{Introduction}
\begin{figure}[t!]
	\begin{center}
		\includegraphics[width=8.5cm,height=4.3cm]{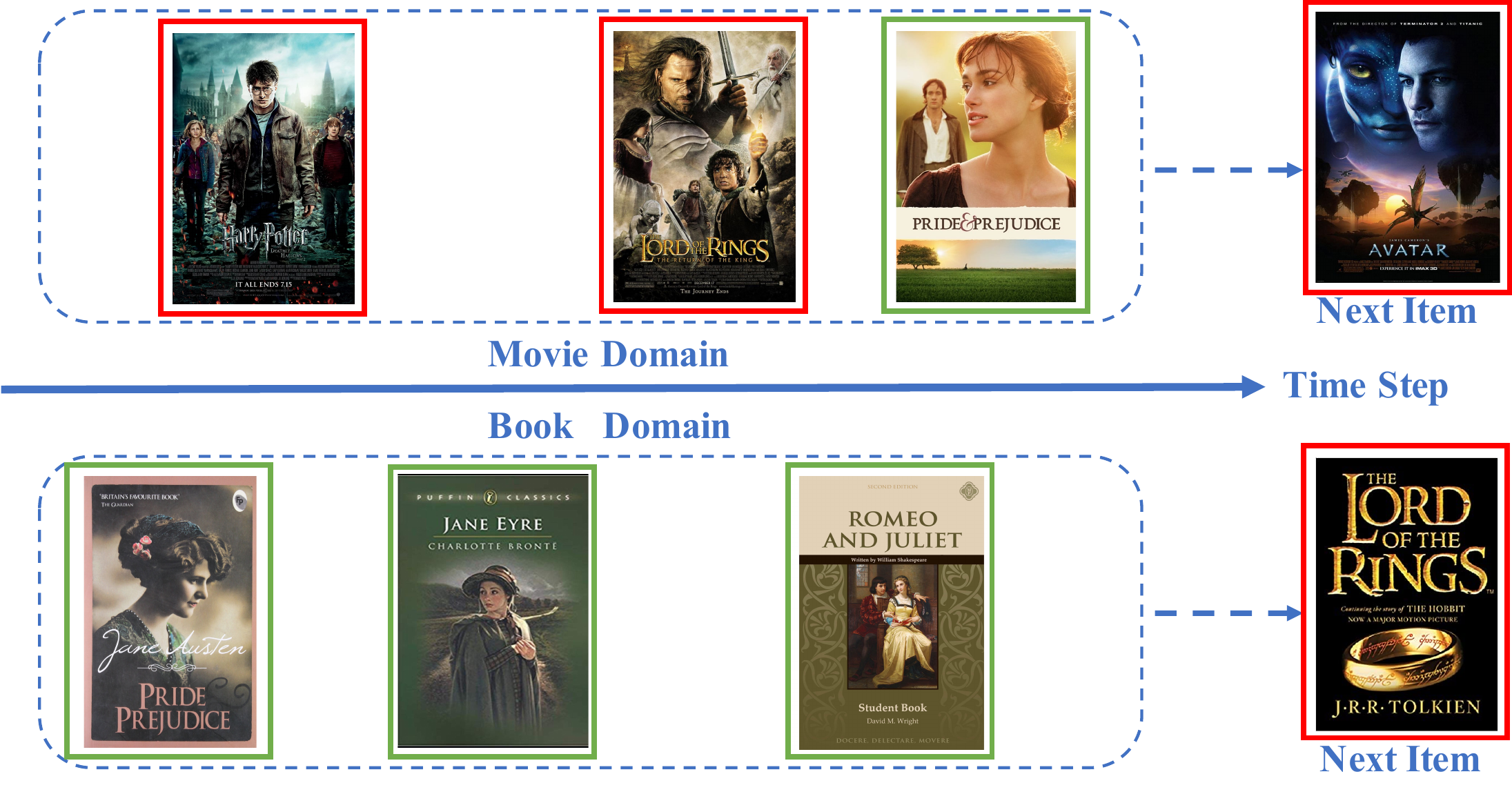}
		\caption{Illustration of user's sequential interactions in \texttt{Movie} and \texttt{Book} domains. Movie or book surrounded by the same color reflects similar user preference, where the green/red represent the ``Romance''/``Fantasy'' preferences.}
		\label{story}
	\end{center}
\end{figure}

In order to model the dynamic user preference, the sequential recommendation (SR), aiming at recommending the next item for a user based on his/her past sequential interactions, has attracted a surge of interest in many web applications such as Amazon and YouTube.
In the real world, however, users usually exhibit partial (or incomplete) preference involved in a specific domain, making the recommendation results practically biased on the observed single-domain historical interactions.

To alleviate the above issue, cross-domain sequential recommendation (CDSR) has been proposed, which attempts to improve multiple domain sequential recommendation performances simultaneously by leveraging rich information from other relative domains. 
Figure \ref{story} gives a toy example of a typical user with historical interactions of two domains: \texttt{Movie} and \texttt{Book} domain.
As shown in the figure, the user recently watched the movie \textit{Lord Rings}, and then he/she appreciates the corresponding book \textit{Lord Rings} as the following interaction.
Intuitively, the recommender system is unlikely to recommend the book \textit{Lord Rings} individually based on the observed book interactions, since there are no explicitly similar interactions reflecting the \textit{Lord Rings} preference in the single \texttt{Book} domain.  
This is because that the users' historical interactions in each single domain may only reflect the partial preference, easily making the recommender system biased on the incomplete (or single-domain) preference.
Fortunately, by taking an overview of the both domains, we can derive the relatively complete (or cross-domain) preference of the current user, namely `Romance' and `Fantasy' interests for both \texttt{Movie} and \texttt{Book} domain, and the `Fantasy' interests may provide positive clues to make right recommendation.
Therefore, it is necessary to capture the precise cross-domain user preference for better unbiased recommendation.

To identify and transfer the valuable clues across domains, a significant paradigm of CDSR research has been proposed by extending the single-domain sequential recommendation methods.
The pioneering work is $\pi$-net~\cite{pinet}, which first generates single-domain representation by modeling item interaction sequence in each single domain, and then transfers the learned single-domain representations into other domains with a gated transferring module.
To enhance the transferring module in $\pi$-net, MIFN~\cite{mifn} further introduces an external knowledge graph transferring module to guide the connection between different domain items.

\begin{figure}[t]
	\begin{center}
		\includegraphics[width=8.5cm,height=6.5cm]{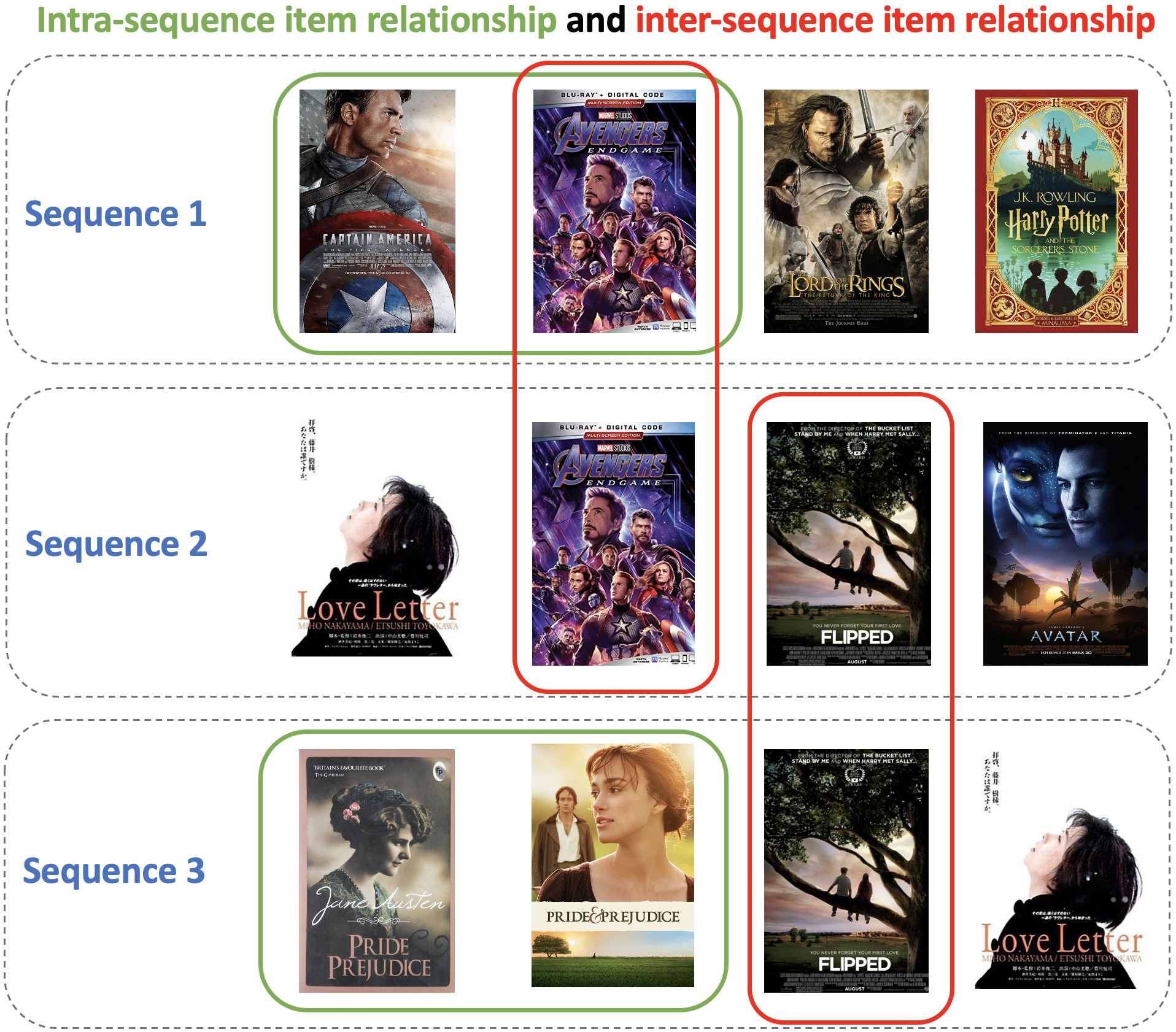}
		\caption{A toy illustration of item relationships. 
		The green boxes reflect the sequential pattern signal of intra-sequence item relationships. The red boxes reflect the collaborative signal of inter-sequence item relationships.}
		\label{intrainter}
	\end{center}
\end{figure}

Despite the promising improvements, the pipeline-style paradigm learns single-domain user preference separately, which usually generates domain-biased user representations.
Simply transferring the biased single-domain preference can be intractable to describe precise cross-domain user preference, which would easily lead to unstable and sub-optimal recommendation results.
Therefore, we argue that it is necessary to \textbf{learn the single- and cross- domain user preference in a joint way} for unbiased information transferring.
More importantly, typical user interactions in different domains usually exhibit related preferences, thus we further consider the \textbf{correlation between the single- and cross- domain user preference.}
Besides, previous CDSR works~\cite{pinet,PSJnet} only focus on modeling the intra-sequence item relationship to capture the sequential pattern signal to obtain sequence representation (i.e, user representation), but ignore the inter-sequence relationship of items (as shown in Figure~\ref{intrainter}), which provides valuable collaborative signal to generate better user representation.
Therefore, we propose to \textbf{capture the intra- and inter- sequence item relationships at same time} for representation learning.

To implement the above idea, we propose a \underline{c}ontrastive \underline{c}ross-\underline{d}omain \underline{s}equential \underline{r}ecommendation model, termed as \textbf{C$^2$DSR}.
We first utilize a graph neural network to mine inter-sequence item collaborative signal, and then exploit a sequential attentive encoder to capture intra-sequence item sequential signal.
Based on the graph and attention modules, we introduce and optimize single- and cross- domain next item prediction objectives to obtain single- and cross- user representations.
Furthermore, we develop a novel contrastive infomax objective to encourage single- and cross- domain representations to be relevant, which leverages the mutual information maximization principle~\cite{infomaxprinciple,dim} to enhance their correlation. 
To validate the effectiveness of our method, we evaluate our model and prior efforts on the Amazon and HVIDEO datasets against previous CDSR works.

Overall, our main contributions are summarized as follows:
\begin{itemize}
	\item  We propose a novel model C$^2$DSR which models the intra- and inter- sequence item relationships to obtain the single- and cross- domain user representations simultaneously.
    
	\item We introduce a novel contrastive infomax objective to encourage the single- and cross- user representations to be relevant by maximizing their mutual information.
	
	\item We conduct extensive experiments on four CDSR datasets, which demonstrates that our C$^2$DSR achieves consistent and significant improvements over previous state-of-the-art baselines. Our source codes and datasets are available at Github\footnote{\url{ https://github.com/cjx96/C2DSR}} for further comparisons.
	
\end{itemize}

\section{Problem Formulation}
In this work, we consider a general CDSR scenario, where each interaction sequence involves two domains, namely domain $X$ and domain $Y$.
Let $\mathcal{S}$ denote the overall interaction sequence set, where each instance $(S^X, S^Y, S)_u \in \mathcal{S}$ belongs to a certain user $u$.
For each instance, the $S^X = [x_1, x_2, \dots, x_{|S^X|}]$ and $S^Y = [y_1, y_2, \dots, y_{|S^Y|}]$ denote the corresponding single-domain interaction sequences, and the $S = [y_1, x_1, x_2, \dots, y_{|S^Y|}, \dots, x_{|S^X|}]$ denotes the cross-domain interaction sequence by merging $S^X$ and $S^Y$ in chronological order, where each $x \in \mathcal{X}$ and $y \in \mathcal{Y}$ are the interacted items, and the $|\cdot |$ denotes the total item number.
Note that $\mathcal{X}$ and $\mathcal{Y}$ denote the entire item set in domain $X$ and domain $Y$, respectively. 
For simplicity, we further introduce three directed item-item matrices $\mathbf{A}^X\in \{0,1\}^{|\mathcal{X}|\times |\mathcal{X}|}, \mathbf{A}^Y\in \{0,1\}^{|\mathcal{Y}|\times |\mathcal{Y}|}, \mathbf{A}\in \{0,1\}^{(|\mathcal{X}|+|\mathcal{Y}|)\times (|\mathcal{X}|+|\mathcal{Y}|)}$ to represent $\mathcal{S}$, where $\mathbf{A}_{ij}^X = 1$ if $x_j$ is the one next item of $x_i$, and $\mathbf{A}_{ij}^X = 0$ otherwise.

Given the observed interaction sequences $(S^X, S^Y, S)_u$, the goal of CDSR is to predict the next item:
\begin{equation}
	\begin{split}
	\mathrm{argmax}_{x_i \in \mathcal{X}} &\mathrm{P}^X \left(x_i|S^X,S^Y,S \right), \text{if } \text{next item} \in \mathcal{X}\\
 	\mathrm{argmax}_{y_j \in \mathcal{Y}} &\mathrm{P}^Y \left(y_j|S^X,S^Y,S \right), \text{if } \text{next item} \in \mathcal{Y}
	\end{split}
	\label{sec_pre_eq:cdsr_goal}
\end{equation}
where $\mathrm{P}^X(x_i|S^X,S^Y,S)\in \mathbb{R}^{|\mathcal{X}|}$ and $\mathrm{P}^Y(y_j|S^X,S^Y,S)\in \mathbb{R}^{|\mathcal{Y}|}$ are the \textbf{probability of the candidate item} in domain $X$ and $Y$, where the highest one is selected as the next recommended item.

\section{methodology}

In this section, we propose our model \textbf{C$^2$DSR}, which captures and transfers valuable information across domains by modeling single-domain and cross-domain representations.
There are three major components of C$^2$DSR: 
(1) Graphical and attentional encoder, which includes an embedding layer, a graph neural network module, and a self-attention module to generate a series of sequential representations (i.e., user representations) for each interaction sequence.
(2) Sequential training objective, which includes two training objectives for single-domain and cross-domain interaction sequences to obtain single-domain and cross-domain user representations.
(3) Contrastive infomax objective, which leverages the mutual information maximization principle to enhance the correlation between single-domain and cross-domain representations.

\subsection{Graphical and Attentional Encoder}
In this section, we introduce the graphical and attentive sequential encoder to capture CDSR data, which contains an embedding layer, a graph neural network module, and a self-attention module.

\subsubsection{Embedding Initialization Layer}

In the embedding mapping stage, to obtain initialized item representations for three single- and cross- item interaction sequences, we introduce three parameter matrices $\mathbf{E}^X \in \mathbb{R}^{|\mathcal{X}| \times d}$, $\mathbf{E}^Y \in \mathbb{R}^{|\mathcal{Y}|\times d}$ and $\mathbf{E} \in \mathbb{R}^{(|\mathcal{X}| + |\mathcal{Y}|)\times d}$, respectively, where $d$ is the dimension of the embeddings.
Besides, to recognize the ordered information of sequence, we define a learnable parameter position embedding matrix $\mathbf{T} \in \mathrm{R}^{M\times d}$ to enhance the input item embeddings for the self-attention module~\cite{bert}, where $M$ is the max length of the interaction sequence, e.g., $M = 30$.

\subsubsection{Graph Neural Network Module}
As a promising way to model the inter-sequence item relationship, we consider employing the graph neural network~\cite{gcn} over all the sequences.
Motivated by recent studies~\cite{lightgcn,sgc}, we also remove the convolution matrix multiplication and non-linear activation function to capture the co-occurrence collaborative filtering signal better. 
Specifically, given the binary directed item-item matrices $\mathbf{A}^X, \mathbf{A}^Y, \mathbf{A}$ and initialized embedding $\mathbf{G}_0^X = \mathbf{E}^X, \mathbf{G}_0^Y = \mathbf{E}^Y, \mathbf{G}_0 = \mathbf{E}$, we have:
\begin{equation}
\small
	\begin{split}
		\mathbf{G}^X_1 = \texttt{Norm}(\mathbf{A}^X)\mathbf{G}^X_0, \quad \mathbf{G}^Y_1 = \texttt{Norm}(\mathbf{A}^Y)\mathbf{G}^Y_0,\quad \mathbf{G}_1 &= \texttt{Norm}(\mathbf{A})\mathbf{G}_0, 
	\end{split}
	\label{sec_math_eq:graph}
\end{equation}
where $\texttt{Norm}(\cdot)$ denote the row-normalized function, $\mathbf{G}_1^X, \mathbf{G}_1^Y, \mathbf{G}_1$ are the convolutional outputs.
By stacking $L$ layers, we could obtain a series outputs $\{\mathbf{G}_0^X,\dots,\mathbf{G}_L^X\}$, $\{\mathbf{G}_0^Y,\dots,\mathbf{G}_L^Y\}$, $\{\mathbf{G}_0,\dots,\mathbf{G}_L\}$.
To fully capture graphical information across layers, we use a $\texttt{Mean}(\cdot)$ function to average them to fulfill item representations as:
\begin{equation}
\small
	\begin{split}
    \mathbf{G}^X = \texttt{Mean}(\mathbf{G}_l^X) + \mathbf{E}^X, \mathbf{G}^Y = \texttt{Mean}(\mathbf{G}_l^Y) + \mathbf{E}^Y,  \mathbf{G} = \texttt{Mean}(\mathbf{G}_l) + \mathbf{E}.
	\end{split}
	\label{sec_math_eq:graph_mean}
\end{equation}

\begin{figure}[t]
	\begin{center}
		\includegraphics[width=8.5cm,height=7cm]{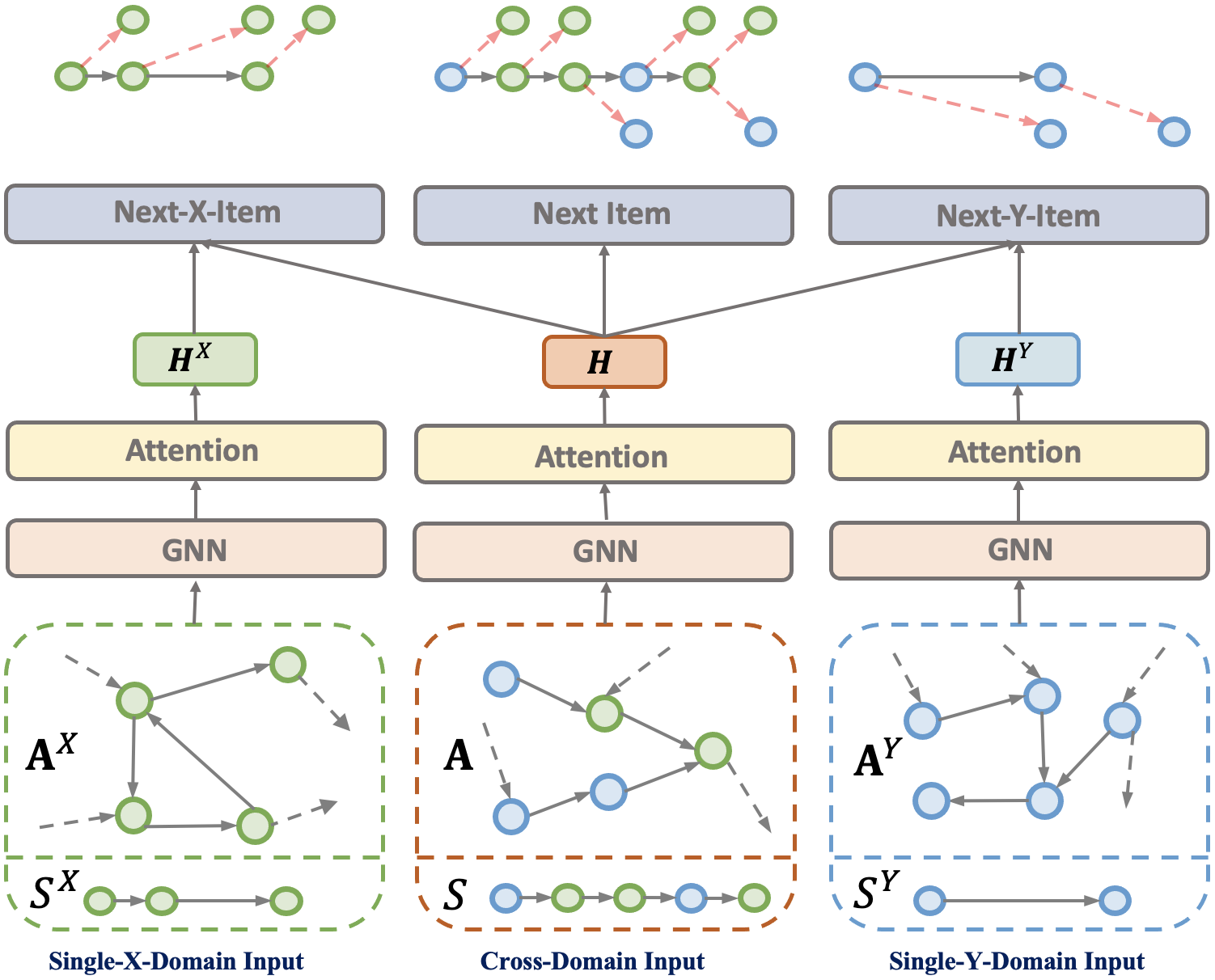}
		\caption{A toy example of sequential training objective for CDSR. The red dotted lines indicate the next prediction item.}
		\label{sec_math_fig:overview}
	\end{center}
\end{figure}

\subsubsection{Self-Attention Module}
To capture the intra-sequence item relationship, we utilize the self-attention module to encode the interaction sequences, since previous SR works prove the superior effectiveness of attention mechanism~\cite{tisasrec,s3rec,bert4rec} against other sequential architectures such as GRU~\cite{gru4rec}.
Similar to SASRec~\cite{sasrec}, in a self-attention module, there are two types of sub-layers: (1) the multi-head self-attention layer captures the complex intra-sequence item dependency in an interaction sequence, (2) the point-wise feed-forward layer endows a non-linearity projection to output the final sequential representations.
For brevity, we employ the padding technique on the input sequences\footnote{We add a \texttt{<pad>} item to the corresponding positions to separate single-domain sequences $S^X$ and $S^Y$ from $S$. For example, $S^X = [\texttt{<pad>}, x_1,x_2,\texttt{<pad>},x_3]$, $S^Y = [y_1, \texttt{<pad>},\texttt{<pad>},y_2,\texttt{<pad>}]$, $S = [y_1, x_1,x_2,y_2,x_3]$, where a constant zero vector $\mathbf{0}$ is used as the embedding for the \texttt{<pad>} item.}, and then formulate the overall encoding process (containing several feed-forward and self-attention layers) as:
\begin{equation}
\small
	\begin{split}
		\bm{H}^X = \mathtt{AttEncoder}^X&(S^X, \mathbf{G}^X),\quad \bm{H}^Y = \mathtt{AttEncoder}^Y(S^Y, \mathbf{G}^Y), \\
		\bm{H} &= \mathtt{AttEncoder}(S, \mathbf{G}), 
	\end{split}
	\label{sec_math_eq:attention}
\end{equation}
where $\bm{H}^X\in \mathbb{R}^{|S|\times d}, \bm{H}^Y \in \mathbb{R}^{|S|\times d}, \bm{H} \in \mathbb{R}^{|S|\times d}$ are sequential outputs, and we adopt several $\mathtt{AttEncoder}$ parameters to encode different interaction sequence $S^X$, $S^Y$ and $S$ for better adaptation.

\subsection{Sequential Training Objective}

Building upon the generated sequential outputs $\bm{H}^X, \bm{H}^Y, \bm{H}$, in this section, we employ two sequential objectives to optimize them to act as single-domain and cross-domain representations (in Figure~\ref{sec_math_fig:overview}).

\subsubsection{Single-Domain Item Prediction}
For the single-domain item prediction, as applied in many sequential recommendation approaches~\cite{sasrec,gru4rec,caser}, the most common strategy is to train the model to recommend the next item based on the observed sequence directly.
Taking the domain $X$ as an example, given single-domain padding interaction sequence $S^X = [\texttt{<pad>},x_1, x_2, \texttt{<pad>}, \dots, x_{t}]$, and its expected next item $x_{t+1}$.
We adopt the commonly used training strategy to optimize our encoder as follows:
\begin{equation}
\small
	\begin{split}
		\mathcal{L}^X_{\text{single}} &= \sum_{S^X \in \mathcal{S}} \sum_{t} \mathcal{L}^{X}_{\text{single}}(S^X, t) \\
		\mathcal{L}^X_{\text{single}}(S^X, t) = -\log& \mathrm{P}^X_{\text{single}}(x_{t+1}| [\texttt{<pad>},x_1, x_2, \texttt{<pad>}, \dots, x_{t}]), 		\\
	\end{split} 
	\label{sec_math_eq:single}
\end{equation}
where the probability $\mathrm{P}^X_{\text{single}}(x_{t+1} | [\texttt{<pad>},x_1, x_2, \texttt{<pad>}, \dots, x_{t}])$ is designed to be proportional to the similarity between all the items $x \in \mathcal{X}$ and the given sequence in the vector space.
Based on the learned representations $\bm{H}^X$ and $\bm{H}$, we calculate single-$X$-domain prediction probability $\mathrm{P}^X_{\text{single}}(\cdot)$ as follows (similarly for domain $Y$):
\begin{equation}
\small
	\begin{split}
		\mathrm{P}^X_{\text{single}}(x_{t+1} | [ \dots, x_{t}]) = \texttt{Softmax}\big(\bm{h}_t^X\mathbf{W}^X + \bm{h}_t\mathbf{W}^X\big)_{x_{t+1}}
	\end{split}
	\label{sec_math_eq:single_prediction}
\end{equation}
where the the $\bm{h}_t^X \in \mathbb{R}^{1\times d}, \bm{h}_t \in \mathbb{R}^{1\times d}$ are sequential representations at position $t$, $\mathbf{W}^X\in \mathbb{R}^{d\times |\mathcal{X}|}$ is the learnable parameter matrix for prediction. 
Specifically, $\texttt{Softmax}\big(\bm{h}_t^X\mathbf{W}^X + \bm{h}_t\mathbf{W}^X\big) \in \mathbb{R}^{|\mathcal{X}|}$ are non-negative prediction probabilities for all items $x \in \mathcal{X}$, and its sum is 1.
 We select the corresponding probability of $x_{t+1}$ from it as our final prediction score.
Note that those formulations (in Eq.(\ref{sec_math_eq:single}-\ref{sec_math_eq:single_prediction})) also holds for single-$Y$-domain as $\mathcal{L}^Y_{\text{single}}$.

In particular, this training strategy could achieve promising results in a single-domain scenario, but it is hard to model the cross-domain interaction sequence.
For example, suppose that we merge the two single-domain sequences to obtain the cross-domain item sequence ``[ \texttt{Movie}, \texttt{Movie}, \texttt{Movie}, \texttt{Movie}] $\rightarrow$ \texttt{Book}''.
Since the observed items are always \texttt{Movie} type, the above training strategy would minimize all \texttt{Book} type probabilities by optimizing $\texttt{Softmax}(\cdot)\in \mathbb{R}^{|\mathcal{X}|+|\mathcal{Y}|}$, even if some \texttt{Book} items show strong relevance with the latest \texttt{Movie} items, which may limit the model effectiveness.
This observation motivates us to devise a more balanced training strategy for cross-domain interaction sequences.

\subsubsection{Cross-Domain Item Prediction}
In a cross-domain interaction sequence, a user usually shows diverse preferences, and the preference in one domain may impact the following interactions in another domain (e.g., watching the movie \textit{Lord Rings} may lead to reading the book \textit{Lord Rings}).
Further, different domains contain different amount interactions, to avoid the prediction result dominated by the data-rich domain, a balanced objective function is required.
That is to say, our model should also preserve the prediction ability in the \texttt{Book} domain when it predicts next the \texttt{Movie} item.
Following this intuition and given next item $x_{t+1}$ or $y_{t+1}$, we define the training strategy for cross-domain sequence $S$ as follows:
\begin{equation}
\small
	\begin{split}
		&\mathcal{L}_{\text{cross}} = \sum_{S \in \mathcal{S}} \sum_{t} \mathcal{L}_{\text{cross}}(S, t), \\
		\mathcal{L}_{\text{cross}}(S, t) = &\left\{
\begin{aligned}
-\log \mathrm{P}^X_{\text{cross}}&(x_{t+1}| [y_1, x_1, x_2, \dots, x_t]), \\
-\log \mathrm{P}^Y_{\text{cross}}&(y_{t+1}| [y_1, x_1, x_2, \dots, x_t]), \\
\end{aligned}
\right.
	\end{split} 
	\label{sec_math_eq:cross}
\end{equation}
where we implement the prediction probability $\mathrm{P}_{\text{cross}}^X(\cdot), \mathrm{P}_{\text{cross}}^Y(\cdot)$ by utilizing the learned representation $\bm{H}$ as follows:
\begin{equation}
\small
	\begin{split}
		\mathrm{P}^X_{\text{cross}}(x_{t+1} | [y_1, x_1, x_2, \dots, x_t]) = \texttt{Softmax}\big(\bm{h}_t\mathbf{W}^X\big)_{x_{t+1}}, \\
		\mathrm{P}^Y_{\text{cross}}(y_{t+1} | [y_1, x_1, x_2, \dots, x_t]) = \texttt{Softmax}\big(\bm{h}_t\mathbf{W}^Y\big)_{y_{t+1}}, \\
	\end{split}
	\label{sec_math_eq:cross_prediction}
\end{equation}
where $\mathbf{W}^X$ is the same parameter matrices in $\mathrm{P}_{\text{single}}^X(\cdot)$.
In the above Eq.(\ref{sec_math_eq:cross}), we decompose the loss function as two equally separate parts for the domain $X$ and $Y$ to optimize, which could preserve the prediction ability for one domain even if the interaction sequence contains many consecutive items in another domain.

Up to now, we have introduced the single- and cross- domain sequential item prediction objective.
As shown in the Figure~\ref{sec_math_fig:overview}, the two type objectives have quite different purposes by leveraging the learned representations.
In detail, $\bm{H}^X$, $\bm{H}^Y$ are only used to predict the next single-domain item, and $\bm{H}$ aims to predict both domains items.
Therefore, by optimizing Eq.(\ref{sec_math_eq:single}) and Eq.(\ref{sec_math_eq:cross}), $\bm{H}^X$, $\bm{H}^Y$ are tended to encode single-domain user preference and $\bm{H}$ is encouraged to act as the cross-domain user preference.

\begin{figure}[t]
 	\begin{center}
 		\includegraphics[width=8cm,height=4.5cm]{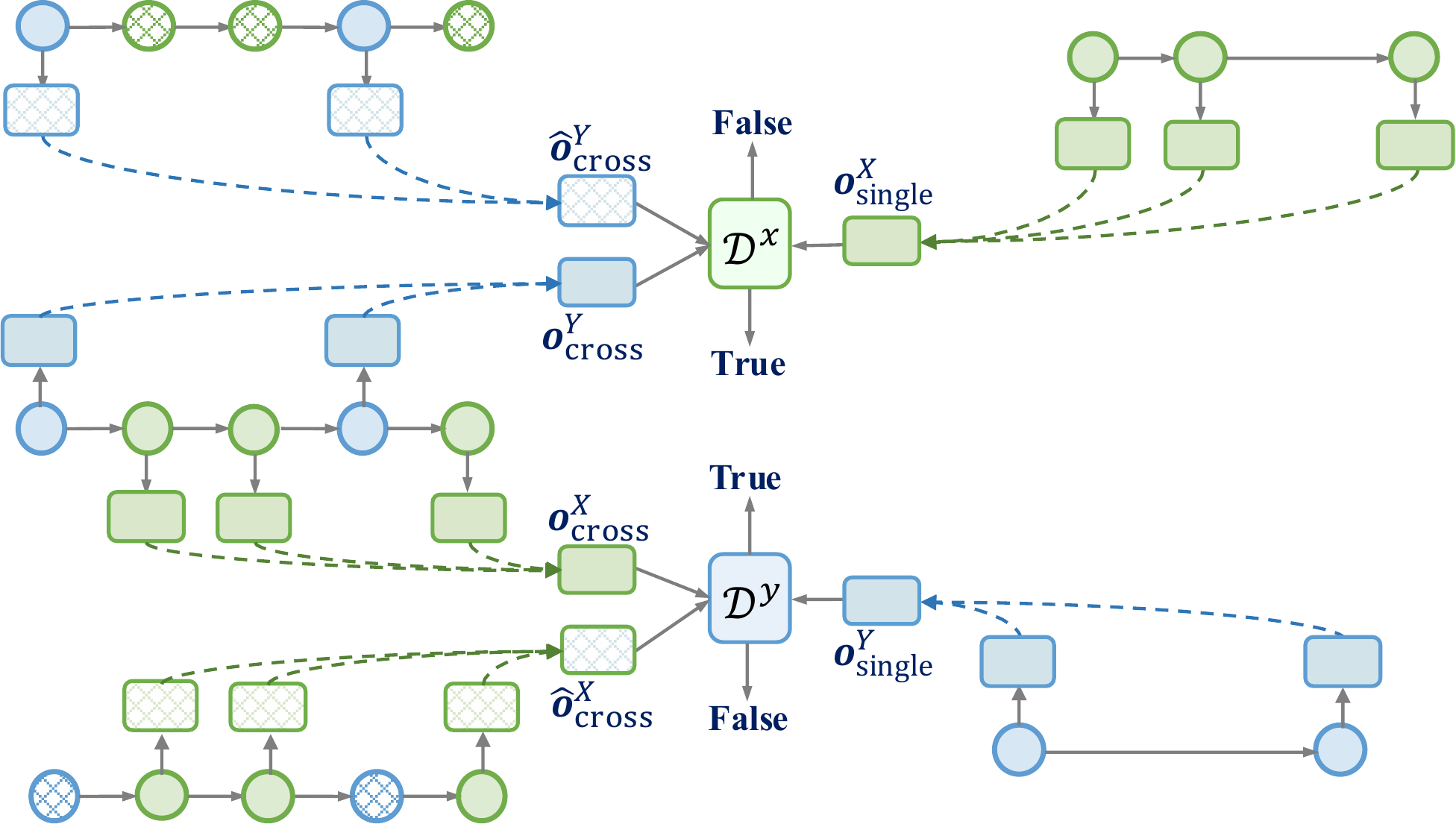}
 		\caption{Illustration of our contrastive infomax. The hollow  green/blue circles denote random items in domain $X$/$Y$.}
 		\label{sec_math_fig:infomaxgraph}
 	\end{center}
\end{figure}
 
\subsection{Contrastive Infomax Objective}

Since the single-domain representations $\bm{H}^X$, $\bm{H}^Y$ are learned from sequence $S^X$, $S^Y$ independently, they tend to derive the partial user's preference in a single domain.
As mentioned before, it can be beneficial to incorporate the cross-domain preference to make better recommendations.
Hence, inspired by the infomax principle~\cite{dgi,bigi}, we develop a novel contrastive infomax objective to improve correlation between the single-domain representations and cross-domain representations.

\subsubsection{Single- and Cross- Domain Prototype Representations}
To represent user's holistic preferences in different interaction sequences $S^X, S^Y, S$, we generate the prototype representations~\cite{fewshot} for them.
Our insight of the prototype representation is borrowed from the prototypical network~\cite{learnfewshot}, which generates prototypes to represent samples of the same type.
We utilize the centroid of sequential representations as to the prototype representation by a simple $\mathtt{Mean}$ operation.
For example, to obtain single-domain prototype representations, the concrete procedures can be formulated as:
\begin{equation}
\small
	\bm{o}^X_{\text{single}} = \mathtt{Mean}(\bm{H}^X), \quad \quad \bm{o}^Y_{\text{single}} = \mathtt{Mean}(\bm{H}^Y),
	\label{sec_math_eq:specificproto}
\end{equation}
where the $\bm{o}^X_{\text{single}}\in \mathbb{R}^{1\times d}$ and $\bm{o}^Y_{\text{single}} \in \mathbb{R}^{1\times d}$ are the $X$-domain and $Y$-domain prototype representations of $S^X$ and $S^Y$, respectively.

For the cross-domain interaction sequence $S$, we generate two cross-domain prototype representations by average corresponding domain item representations:
\begin{equation}
\small
	\begin{split}
	\bm{o}^X_{\text{cross}} = \mathtt{Mean}(\{\bm{h}_t : S_t \in \mathcal{X}\}), \quad
	\bm{o}^Y_{\text{cross}} = \mathtt{Mean}(\{\bm{h}_t : S_t \in \mathcal{Y}\}), \\
	\end{split}
	\label{sec_math_eq:unifiedproto}
\end{equation}
where the $\bm{o}^X_{\text{cross}}$ and $\bm{o}^Y_{\text{cross}}$ are the cross-domain prototype representations for domain $X$ and $Y$, which encode the user's holistic preferences from both domains information.

\subsubsection{Infomax Objective}
After obtaining the prototype representations, we follow the intuition from DIM~\cite{dim} and use a noise-contrastive objective between the samples from joint (positive examples) and the product of marginals (negative examples) to formulate our contrastive infomax objective.
To generate negative samples from the positive cross-domain interaction sequence $S$, we devise two corruption functions $\mathtt{Corrupt}^X$ and $\mathtt{Corrupt}^Y$ with the negative sampling trick for domain $X$ and $Y$ respectively.
The detail of our corruption functions can be performed as:
\begin{equation}
\small
	\begin{split}
	\widehat{S}^X &= \mathtt{Corrupt}^X(S) = [\widehat{y}_1, x_1, x_2, \widehat{y}_2, \dots], \\
	\widehat{S}^Y &= \mathtt{Corrupt}^Y(S) = [y_1, \widehat{x}_1, \widehat{x}_2, y_2, \dots],
	\label{sec_math_eq:corruption}
	\end{split}
\end{equation}
where $\widehat{x}$ and $\widehat{y}$ are randomly selected items in corresponding domain, and $\widehat{S}^X$ and $\widehat{S}^Y$ are the corrupted cross-domain interaction sequences.
By Eq.(\ref{sec_math_eq:attention}) and Eq.(\ref{sec_math_eq:unifiedproto}), we produce the prototype representations $\widehat{\bm{o}}^X_{\text{cross}}$ and $\widehat{\bm{o}}^Y_{\text{cross}}$ for $\widehat{S}^X$ and $\widehat{S}^Y$ respectively.

Considering the user may have correlated preference between two domains, we hope the prototype representations from two domains are inherently relevant.
Taking domain $X$ as an example, we hope single-$X$-domain representation $\bm{o}^X_{\text{single}}$ is relevant to $\bm{o}^Y_{\text{cross}}$ with true $X$ domain items, but is irrelevant to $\widehat{\bm{o}}^Y_{\text{cross}}$ with false $X$ domain items.
Such a design not only enforces correlations between two domains, but also makes cross-domain representations (e.g., $\bm{o}^Y_{\text{cross}}$) focusing on the real sequential interaction in the domain $X$. 
Thereby, our infomax objective $\mathcal{L}_{\text{disc}}^X$ is defined as (note that also holds for domain $Y$ and  leads to $\mathcal{L}_{\text{disc}}^Y$):
\begin{equation}
\small
	\begin{split}
		\mathcal{L}_{\text{disc}}^X = \!\!\!\!\!\!\!\!    \sum_{(S^X, S^Y, S)_u \in \mathcal{S}} \!\!\!\!\!\!\! - \Big(\!\log \mathcal{D}^X(\bm{o}^X_{\text{single}},\bm{o}^Y_{\text{cross}}) \! + \! \log\big(1 - \mathcal{D}^X(\bm{o}^X_{\text{single}},\widehat{\bm{o}}^Y_{\text{cross}})\big)\Big)
	\end{split} 
	\label{sec_math_eq:infomax}
\end{equation}
where the $\mathcal{D}_X$ can be regarded as a binary discriminator to measure single- and cross- domain prototype representation pairs by a bilinear mapping function:
\begin{equation}
\small
	\mathcal{D}^X(\bm{o}^X_{\text{single}},\bm{o}^Y_{\text{cross}}) = \sigma\big(\bm{o}^X_{\text{single}}\mathbf{W}^X_{\text{disc}}(\bm{o}^Y_{\text{cross}})^\top\big),
	\label{sec_math_eq:discri}
\end{equation}
where $\mathbf{W}^X_{\text{disc}} \in \mathbb{R}^{d \times d}$ is a learnable parameter matrix, $\sigma$ is the $\mathtt{Sigmoid}(\cdot)$ function. As discussed in previous works~\cite{dgi,bigi}, the binary cross-entropy loss in Eq.(\ref{sec_math_eq:infomax}) is an effective mutual information (MI) estimator. 
It can maximize the MI between the single- and cross- domain representations, and capture the correlation between them, since it measures the Jensen–Shannon (JS) divergence between the joint distribution and the product of marginals.
Further, it follows the min-max objective as formalized in generative adversarial network (GAN)~\cite{gan}, and the GAN objective is closely related to JS divergence that can be used in MI estimation.

\subsection{Model Training and Evaluation}
The total loss function of C$^2$DSR contains two types:
\begin{equation}
\small
	\mathcal{L} = \underbrace{\lambda(\mathcal{L}_{\text{cross}} + \mathcal{L}^X_{\text{single}} + \mathcal{L}^Y_{\text{single}})}_{\text{Sequential training objective}} \quad+ \underbrace{(1 - \lambda)(\mathcal{L}^X_{\text{disc}} + \mathcal{L}^Y_{\text{disc}})}_{\text{Contrastive infomax objective}}
	\label{sec_math_eq:loss}
\end{equation}
where $\lambda$ is the harmonic factor. Our C$^2$DSR can be optimized with the mini-batch manner~\cite{adam}, and keeps the similar time complexity\footnote{Empirically, in the same running environment, C$^2$DSR and SASRec would cost around 54s and 40s per epoch on ``Food-Kitchen'' dataset for training on a Tesla T4 GPU. Details are reported in implementation setting section.} with other attention-architecture SR approaches. 

In the evaluation stage, we utilize the corresponding single-domain and cross-domain representations to make prediction.
For instance, given the latest representations $\bm{h}^X_{|S|}$ and $\bm{h}_{|S|}$, we select the item with highest prediction score in domain $X$ as the recommended next item (see Section~\ref{sec:leave_one_out}):
\begin{equation}
\small
	\begin{split}
	\mathrm{argmax}_{x_i \in \mathcal{X}} &\mathrm{P}^X \left(x_i|S^X,S^Y,S\right), \quad \text{where} \\
	\mathrm{P}^X(x_i | S^X,S^Y,S) &= \texttt{Softmax}\big(\bm{h}_{|S|}^X\mathbf{W}^X + \bm{h}_{|S|}\mathbf{W}^X\big)_{x_i}. \\
	\end{split}
	\label{sec_math_eq:predict}
\end{equation}
The prediction function also holds for domain $Y$, achieving the final task goal in Eq.~(\ref{sec_pre_eq:cdsr_goal}).

\section{Experiments}
In this section, we evaluate the performance of C$^2$DSR in comparison with various approaches.
Moreover, we conduct detailed studies about model variants to show the effectiveness of our model components and analyze the impact of hyperparameters.

\subsection{Datasets}
As used in many cross-domain recommendation methods, we also utilize the public available Amazon\footnote{\url{http://jmcauley.ucsd.edu/data/amazon/index_2014.html}} (E-commerce platform) and HVIDEO\footnote{\url{https://bitbucket.org/Catherine_Ma/pinet_sigir2019/src/master/HVIDEO/}} (TV service platform) datasets to build the CDSR scenarios.
Following previous works~\cite{pinet, mifn}, we select the following six domains to generate three CDSR scenarios for experiments: ``Food-Kitchen'' (Amazon), ``Movie-Book'' (Amazon) and ``Entertainment-Education'' (HVIDEO).
For fair comparisons with previous methods, we first extract users who have interactions in both domains and then filter out some users and items that the number of interactions are fewer than 10.
Furthermore, to satisfy the sequential constraints, we preserve those cross-domain interaction sequences containing at least 3 items from each domain within a period of time (e.g., a month for the ``Movie-Book'' and a year for the ``Food-Kitchen'').
In the training/validation/test partition, we are different from prior works which randomly select samples to construct the training/validation/test datasets resulting in the potential information leak problem.
We analyze the improper setting by calculating those users' interaction sequences in validation/test sets that happened before than corresponding users' interaction sequences in training set (using future interaction sequences to predict the past interactions), and we find about 60\% validation/test samples were shown in preprocessed Amazon datasets.
In our work, to avoid this problem, the users' lastest interaction sequences are equally divided into the validation/test set, and the other interaction sequences for training set.
Statistics of our corrected datasets in CDSR scenarios are summarized in Table \ref{sec_exp_tab:dataset}.

\begin{table}[t]
\centering
\caption{Statistics of Three CDSR scenarios.}
\setlength{\tabcolsep}{5pt}
\resizebox{8.5cm}{!}{
\begin{tabular}{cccccc}
\toprule
\textbf{Scenarios}  &\textbf{\#Items} &\textbf{\#Train} &\textbf{\#Valid}  &\textbf{\#Test} &\textbf{Avg.length}\\ \midrule
Food  &29,207  &\multirow{2}{*}{34,117} &2,722   &2,747   &\multirow{2}{*}{9.91}     \\ 
Kitchen  &34,886 &\multirow{2}{*}{}  &5,451   &5,659  &\multirow{2}{*}{}     \\
\midrule
Movie  &36,845  &\multirow{2}{*}{58,515} &2,032  &1,978  &\multirow{2}{*}{11.98}    \\ 
Book  &63,937 &\multirow{2}{*}{}  &5,612  &5,730 &\multirow{2}{*}{}   \\ 
\midrule
Entertainment  &8,367  &\multirow{2}{*}{120,635} &4,525  &4,485  &\multirow{2}{*}{29.94}    \\ 
Education  &11,404 &\multirow{2}{*}{}  &2,404  &2,300 &\multirow{2}{*}{}   \\  
\bottomrule
\end{tabular}
}
\label{sec_exp_tab:dataset}
\end{table}

\begin{table*}[h!]
\footnotesize
\caption{Experimental results (\%) on the Food-Kitchen scenario.}
\label{foodkitchen}
\setlength\tabcolsep{11.5pt}{
\begin{tabular}{lcccccccccccc}
\toprule
\multirow{3}{*}{\bf Methods} & \multicolumn{6}{c}{\bf Food-domain recommendation} & \multicolumn{6}{c}{\bf Kitchen-domain recommendation}     \\
\cmidrule(r){2-7}\cmidrule{8-13}&
\multirow{2}{*}{MRR} &\multicolumn{2}{c}{NDCG} & \multicolumn{3}{c}{HR} & \multirow{2}{*}{MRR} &\multicolumn{2}{c}{NDCG} & \multicolumn{3}{c}{HR}\\
\cmidrule(r){3-4}\cmidrule(r){5-7}\cmidrule(r){9-10}\cmidrule{11-13} &  & @5 & @10  & @1  & @5  & @10  &  & @5  & @10  & @1  & @5  & @10   \\
\midrule
BPRMF  &  4.10  &   3.55   &  4.03 &   2.42 &  4.51  &   5.95 &
 2.01 &  1.45  &  1.85 &  0.73 &  2.18 &   3.43  \\
ItemKNN    &    3.92  &   3.51   &  3.97 &   2.41 &  4.59  &   5.98 &
 1.89 &  1.28  &  1.75 &  0.58 &  1.99 &   3.26  \\
\midrule
NCF-MLP &   4.49  &   3.94   &  4.51 &   2.68 &  5.10  &   6.86 &
 2.18 &  1.57  &  2.03 &  0.91 &  2.23 &   3.65  \\

CoNet    &   4.13  &   3.61   &  4.14 &   2.42 &  4.77  &   6.35 &
 2.17 &  1.50  &  2.11 &  0.95 &  2.07 &   3.71  \\
\midrule
GRU4Rec   &   5.79  &   5.48   &  6.13 &   3.63 &  7.12  &   9.11 &
 3.06 &  2.55  &  3.10 &  1.61 &  3.50 &   5.22  \\
SASRec   &   7.30  &   6.90   &  7.79 &   4.73 &  8.92  &   11.68  &
 3.79 &  3.35  &  3.93 &  1.92 &  4.78 &   6.62  \\ 
SR-GNN   &    7.84  &   7.58  & 8.35   &   5.03 &  9.88   &   12.27  &
  4.01 &   3.47   &  4.13 &  2.07 &  4.80 &   6.84  \\

\midrule
$\pi$-Net  & 7.68  &   7.32  & 8.13   &   5.25 &  9.25   &  11.75  &
  3.53 &   2.98  &  3.73 &  1.57 &  4.34 &   6.67  \\
PSJNet  & 8.33  &   8.07  & 8.77   &   5.73 &  10.28   &  12.45  &
  \underline{4.10} &   \underline{3.68}   &  \underline{4.32} &  2.14 &  \underline{5.17} &   \underline{7.15}  \\
MIFN  & \underline{8.55}  &   \underline{8.28}  & \underline{9.01}   &   \underline{6.02} &  \underline{10.43}   &  \underline{12.71}  &
  4.09 &   3.57   &  4.29 &  \underline{2.21} &  4.86 &   7.08  \\
\midrule
C$^2$DSR &\textbf{8.91} &\textbf{8.65} &\textbf{9.71} &\textbf{5.84} &\textbf{11.24} &\textbf{14.54} 
&\textbf{4.65} &\textbf{4.16} &\textbf{4.94} &\textbf{2.51} &\textbf{5.74} &\textbf{8.18}\\
\bottomrule
\end{tabular}
}
\end{table*}

\begin{table*}[t]
\footnotesize
\caption{Experimental results (\%) on the Movie-Book scenario. 
}
\label{moviebook}
\setlength\tabcolsep{11.5pt}{
\begin{tabular}{lcccccccccccc}
\toprule
\multirow{3}{*}{\bf Methods} & \multicolumn{6}{c}{\bf Movie-domain recommendation} & \multicolumn{6}{c}{\bf Book-domain recommendation}     \\
\cmidrule(r){2-7}\cmidrule{8-13}&
\multirow{2}{*}{MRR} &\multicolumn{2}{c}{NDCG} & \multicolumn{3}{c}{HR} & \multirow{2}{*}{MRR} &\multicolumn{2}{c}{NDCG} & \multicolumn{3}{c}{HR}\\
\cmidrule(r){3-4}\cmidrule(r){5-7}\cmidrule(r){9-10}\cmidrule{11-13} &  & @5 & @10  & @1  & @5  & @10  &  & @5  & @10  & @1  & @5  & @10   \\
\midrule
BPRMF  &  2.96 &   2.18    &   2.80    &    1.41 &   3.03  &   4.95 & 
 1.27   &   0.85 &  1.17   &    0.48  &    1.23    &    2.25  \\

ItemKNN    &  2.92 &   2.17    &   2.88    &    1.26 &   3.13  &   5.35 & 
 1.26   &   0.86 &  1.10   &    0.48  &    1.25    &    2.00  \\
\midrule
NCF-MLP &  3.05 &   2.26    &   2.96    &    1.41 &   3.13  &   5.30 & 
 1.43   &   1.06 &  1.26   &    0.62  &    1.39    &    2.18  \\
CoNet   &    3.07 &   2.42    &   3.01    &    1.31 &   3.48  &   5.35 & 
 1.45   &   1.04 &  1.28   &    0.64  &    1.44    &    2.19  \\
\midrule
GRU4Rec   &  3.83 &   3.14 &  3.73 &  2.27 &  3.39  &  5.40  &
 1.68 &  1.34   &  1.52 &  0.91 &  1.81 &  2.37  \\

SASRec   &   3.79 &   3.23 &  3.69 &  2.37 &  3.99  &  5.20  &
 1.81 &  1.41   &  1.71 &  0.95 &  1.83 &  2.75  \\

SR-GNN   &  3.85 &  3.27   &  3.78 &  2.22 &  4.19 &  5.81 &
  1.78 &  1.40   &  1.66 &  0.89 &  1.90 &  2.72  \\

\midrule
$\pi$-Net  &   4.16  &   3.72  & 4.17   &   2.52 & 4.75   &   6.11  
& 2.17 &   1.84   &  2.03 &  1.43 &  2.25 &   2.84  \\
PSJNet  &  4.63 &  4.06 &   4.76 &  2.78  & 5.30  &  7.53 & 
 2.44  &  2.07 &   \underline{2.35} &   \underline{1.66}  &  \underline{2.58} &  \underline{3.28} \\

MIFN  &  \underline{5.05} &  \underline{4.21} &   \underline{5.20} &  \underline{2.83}  & \underline{5.51}  &  \underline{8.29} &  \underline{2.51}  &  \underline{2.12} &   2.31 &   1.60  &  2.46 &  3.07 \\

\midrule
C$^2$DSR &\textbf{5.54}  &\textbf{4.76} &\textbf{5.76} &\textbf{3.13} &\textbf{6.47} &\textbf{9.55} 
&\textbf{2.55} &\textbf{2.17} &\textbf{2.45} &\textbf{1.71} &\textbf{2.84} &\textbf{3.75}\\
\bottomrule
\end{tabular}
}
\end{table*}

\begin{table*}[h!]
\footnotesize
\caption{Experimental results (\%) on the Entertainment-Education scenario.}
\label{hvideo}
\setlength\tabcolsep{10pt}{
\begin{tabular}{lcccccccccccc}
\toprule
\multirow{3}{*}{\bf Methods} & \multicolumn{6}{c}{\bf Entertainment-domain recommendation} & \multicolumn{6}{c}{\bf Education-domain recommendation}     \\
\cmidrule(r){2-7}\cmidrule{8-13}&
\multirow{2}{*}{MRR} &\multicolumn{2}{c}{NDCG} & \multicolumn{3}{c}{HR} & \multirow{2}{*}{MRR} &\multicolumn{2}{c}{NDCG} & \multicolumn{3}{c}{HR}\\
\cmidrule(r){3-4}\cmidrule(r){5-7}\cmidrule(r){9-10}\cmidrule{11-13} &  & @5 & @10  & @1  & @5  & @10  &  & @5  & @10  & @1  & @5  & @10   \\
\midrule
BPRMF  & 45.97  &   47.38   &  50.11 &   35.98 &  57.65  &   66.08 &
 46.50 &  47.51  &  49.27 &  38.69 &  54.86 &   60.26  \\
ItemKNN    & 47.81  &   49.24   &  52.01 &   37.91 &  59.44  &   67.93 &
 46.22 &  47.23  &  49.22 &  38.08 &  54.86 &   61.04  \\
\midrule
NCF-MLP &   44.94  &   47.57   &  50.61 &   31.66 &  61.73  &   71.08 &
 46.24 &  48.73  &  50.66 &  35.17 &  60.56 &   66.43  \\

CoNet    &   45.76  &   48.31   &  51.63 &   32.30 &  62.45  &   72.68 &
 47.83 &  50.11  &  52.22 &  36.60 &  61.65 &   68.04  \\
\midrule
GRU4Rec   &   45.61  &   47.48   &  51.46 &   32.73 &  61.06  &   73.35 &
 51.35 &  53.88  &  56.21 &  39.95 &  66.13 &   73.21  \\
SASRec   &   50.44  &   52.67   &  56.10 &   37.05 &  66.39  &   76.92  &
 53.69 &  55.87  &  58.64 &  41.86 &  68.04 &   76.56  \\ 
SR-GNN   &    50.67  &   52.77  & 56.47   &   37.43 &  66.30   &   \underline{77.17}  &
  54.74 &   56.73   &  59.69 &  43.08 &  68.43 &   \underline{77.47}  \\

\midrule
$\pi$-Net  & 52.68  &   54.88  & 57.63   &   \underline{40.91} &  67.15   &  75.11  &
  55.05 &   57.23   &  59.32 &  44.26 &  68.21 &   74.65  \\
PSJNet  & \underline{53.50}  &   \textbf{57.57}  & \textbf{60.07}   &   \textbf{42.52} &  \underline{68.94}   &  76.56  & \underline{55.94} &   \underline{58.18}  & \underline{60.15} &  \textbf{45.21} &  \underline{69.01} &   75.08  \\
\midrule
C$^2$DSR &\textbf{53.87} &\underline{56.20} &\underline{59.35} &40.89 &\textbf{69.45} &\textbf{79.08} 
&\textbf{56.72} &\textbf{59.05} &\textbf{61.56} &\underline{45.13} &\textbf{71.04} &\textbf{79.21}\\

\bottomrule
\end{tabular}
}
\end{table*}

\subsection{Experimental Setting}
\subsubsection{Evaluation Protocol} 
\label{sec:leave_one_out}
Following previous works~\cite{sasrec,bert4rec,s3rec}, we also leverage the leave-one-out method to calculate the recommendation performance.
To guarantee unbiased evaluation, we follow Rendle’s literature~\cite{kddmetric} to calculate 1,000 scores for each validation/test case (including 999 negative items and 1 positive item).
Then, we report the \textit{Top-K} recommendation performance of the 1,000 ranking list in terms of MRR (Mean Reciprocal Rank)~\cite{mrr}, NDCG@\{5, 10\} (Normalized Discounted Cumulative Gain)~\cite{ndcg} and HR@\{1, 5, 10\} (Hit Ratio).

\subsubsection{Compared Baselines} 
\textbf{In this section, we explain four classes of baselines and their key operations for adaptation}.

	\textit{Traditional recommendation baselines}: (1) \textbf{BPRMF}~\cite{bprmf} is a well-known method which devises a pairwise ranking loss function to learn users and items representations. (2) \textbf{ItemKNN}~\cite{itemknn} follows the metric learning idea, which assumes those interacted items of a user have higher similarity than the items not interacted yet. For those methods, we ignore sequential constraints of user-item interaction sequences and train them on mixed user-item pair datasets. 
	
	\textit{Cross-domain recommendation baselines}: (1) \textbf{NCF-MLP}~\cite{neumf} is a famous approach which learns representations by MLP networks. To adapt two domains, we employ it to learn user/item representations by two separate base MLP networks with a shared initialized user embedding layer. (2) \textbf{CoNet}~\cite{conet} is a classic method of cross-domain recommendation, which first models interactions of two domains by two base networks, and then transfers information by a cross-network between the two base networks. For them, we also ignore sequential constraints of user-item interaction sequences. 
	
	\textit{Sequential recommendation baselines}: (1) \textbf{GRU4Rec}~\cite{gru4rec} applies GRU-achitecture to model interaction sequence for SR. (2) \textbf{SASRec}~\cite{sasrec} is one of the state-of-the-art baselines for SR, which uses the  self-attention mechanism to model the interaction sequence. (3) \textbf{SR-GNN}~\cite{srgnn} is a pioneer work to apply graph neural network to capture high-order relationships between items for SR. For those methods, we preserve the sequential constraints and train them on mixed interaction sequence datasets directly.

	\textit{Cross-domain sequential recommendation baselines}: (1) \textbf{$\pi$-Net}~\cite{pinet} is the pioneering work for CDSR, which devises a novel gating recurrent module to model and transfer knowledge across different domains. (2) \textbf{PSJNet}~\cite{PSJnet} is extended from $\pi$-Net, which introduces a parallel split-join scheme to transfer the different user intention across domains. (3) \textbf{MIFN}~\cite{mifn} is recently proposed approach for CDSR, which constructs a knowledge graph to guide the connection between items from other domains to transfer information across domains.  For those methods, we use the \textbf{self-attention module} as their sequence encoder.

\begin{figure*}[t]
	\begin{center}
		\includegraphics[width=17cm,height=3.5cm]{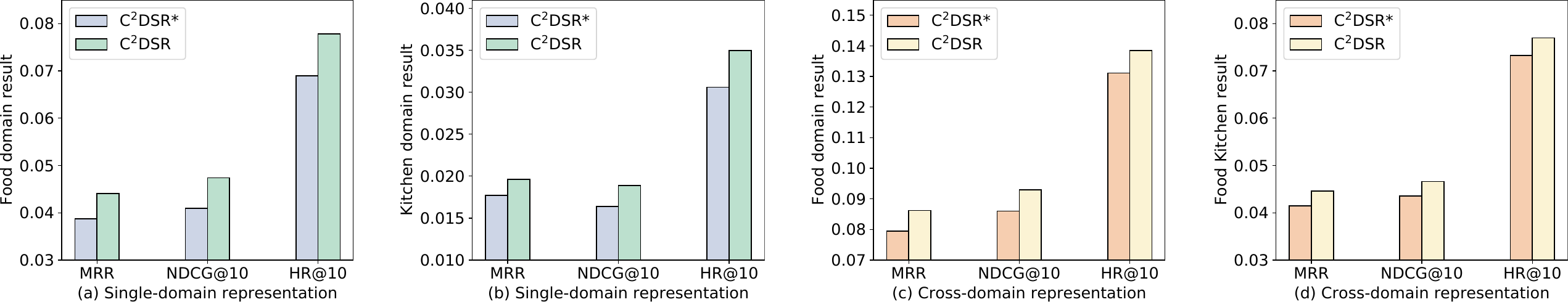}
		\caption{The predictive results of single-domain and cross-domain representations on Food-Kitchen.}
		\label{infomaxanalysis}
	\end{center}
\end{figure*}
\subsubsection{Implementation and Hyperparameter Setting}
For fair comparisons, we keep the same hyperparameter settings as MIFN.
For all methods, we first fix the embedding size $d$ and mini-batch batch size $B$ as 256, the dropout rate is fixed as 0.3, the $L_2$ regularizer coefficient is selected from $\{0.0001$, 0.00005, $0.00001\}$, the learning rate is selected from $\{0.001$, 0.0005, $0.0001\}$, the training epoch is fixed as 100 to get the best result, the harmonic factor $\lambda$ is selected from 0.1 to 0.9 with step length 0.1, the deep of GNN $L$ is selected from $\{1,2,3,4\}$, and the Adam~\cite{adam} optimizer is used to update all parameters. 
Besides, as recommended in SASRec, we adapt two single-head attention blocks and the learned positional embedding in this work, and the channel number of $\pi$-net and PSJNet is 5 as suggested in their literature.
We select the best evaluation results according to the highest MRR performance on the validation set.

\subsection{Performance Comparisons}
Table~\ref{foodkitchen}, \ref{moviebook} and \ref{hvideo} show the performance of compared methods on the ``Food-Kitchen'', ``Movie-Book'' and ``Entertainment-Education'' CDSR scenarios. 
Note that we do not provide MIFN results on ``Entertainment-Education'', since we could not obtain the item knowledge graph on the HVIDEO datasets.
The best performance is in boldface and the second is underlined. 
From them, we have several insightful observations:
(1) For the cross-domain recommendation baselines, CoNet and NCF-MLP show better performance than traditional baselines BPRMF and ItemKNN, which validates that considering the difference between domains and transferring such knowledge across domains could be helpful for better domain adaptation. 
(2) For the SR baselines, GRU4Rec, SASRec and SR-GNN show promising performances in the mixed sequential recommendation datasets. Meanwhile, those methods significantly outperform BPRMF, ItemKNN, NCF-MLP and CoNet. The reason is that capturing the sequential characteristics of interactions could provide valuable clues to make precise recommendations.
Moreover, it is obvious that SASRec and SR-GNN achieve robust performance than GRU4Rec in the ``Entertainment-Education'' CDSR scenario.
After our statistics, we think the reason might be the sequence length, this scenario has much longer sequence than others (as shown in Table~\ref{sec_exp_tab:dataset}), and the self-attention encoder has stronger ability to model the longer sequence than GRU.
(3) For the CDSR baselines, $\pi$-Net, PSJNet and MIFN reach superior performance to other baselines, which indicates the cross-domain information is also beneficial to enhance recommendation performance. 
Besides, those methods show distinct prediction performances, which indicates the transfer strategy is vital to model the cross-domain user preference.
(4) Our C$^2$DSR largely outperforms all baselines in many metrics, proving the high effectiveness of our model in CDSR task. 
This fact indicates that utilizing the inter- and intra- sequence item relationships to model the single- and cross- domain user preference is important for CDSR. 
Compared with previous works ($\pi$-Net, PSJNet), it seems that using an explicit way to guide the information transferring is more useful than modeling implicitly transferring module directly, such as using the cross-domain item knowledge graph of MIFN, our proposed sequential objectives and contrastive infomax objective.
(5) An interesting phenomenon is that the results on  ``Entertainment-Education'' are much higher than others.
This is because that users sometimes conduct repeat behaviors~\cite{repeatnet} in TV platform, e.g., watching the same teleplay continuously during a period of time, making it easy to make recommendation.

\subsection{Discussion of Model Variants}

This section investigates the effectiveness of our model components and the extensibility of the proposed contrastive infomax objective.
We conduct several model variants, and the ``Food-Kitchen'' and ``Entertainment-Education'' results of these model variants are provided in Table~\ref{ablation}, respectively.
Specifically, we use ``++'' to denote that the model is incorporated with our graph neural network and trained by our proposed cross-domain sequential training objective.
We use ``$*$'' to denote that the model is trained without our proposed infomax objective.
Besides, C$^2$DSR(GRU4Rec) adopts GRU4Rec as the sequential encoder but with the same infomax objective.

From Table~\ref{ablation}, we have several insightful observations: (1) The ``++'' marked methods show significant improvements over the original methods, which demonstrates that modeling the inter-sequence and intra-sequence item relations and training with our proposed cross-domain objective is beneficial to fit the cross-domain interaction sequence.
(2) The ``$*$'' marked methods consistently outperform than the ``++'' counterparts, which reveals that jointly modeling the single-domain and cross-domain user preferences is also critical for CDSR task.
(3) Compared with other attention based baselines, the performance of model variant C$^2$DSR(GRU4Rec) is also satisfying, which indicates that our proposed infomax objective can be seamlessly incorporated into other sequential encoders to enhance their effectiveness by capturing the correlation between single- and cross- domain user representations.

\subsection{Discussion of the Infomax Impact}

In this section, to further validate that our presented contrastive infomax objective is able to enhance the single-domain and cross-domain representations, we conduct an analysis between C$^2$DSR and C$^2$DSR$^*$ (trained without the infomax objective). 
As shown in Figure~\ref{infomaxanalysis}, we only leverage the single-domain representation or the cross-domain representation to test the predictive power in each domain.
From it, we can observe that: (1) The cross-domain representation has more powerful predictive ability than the single-domain representation, which indicates that it is vital to consider other relevant domain to enhance each single-domain sequential recommendation.
(2) Compared with C$^2$DSR$^*$, our proposed infomax objective could enhance the prediction ability of single-domain and cross-domain representations simultaneously.
We assume the reason is that our contrastive infomax objective could enable the single-domain representation to perceive the complete user preferences and encourage the cross-domain representation to provide more general preferences in both domains.

\begin{table}[t]
\centering
\caption{Variants results on Food-Kitchen.
}
\label{ablation}
\resizebox{8.5cm}{!}{
\begin{tabular}{lcccccc}
\toprule
\multirow{2}{*}{\bf Model Variants} & \multicolumn{3}{c}{\bf Food-domain} & \multicolumn{3}{c}{\bf Kitchen-domain}     \\
\cmidrule(r){2-4}\cmidrule{5-7}
& MRR  & NDCG@10  & HR@10 & MRR  & NDCG@10  & HR@10   \\

\midrule
GRU4Rec  &   5.79 &  6.13  &   9.11 &  3.06 &  3.10 &   5.22 \\
GRU4Rec++  &  6.92  &   7.18   &  10.10 &   3.21 &  3.28  &   5.63 \\
C$^2$DSR(GRU4Rec)$^*$ &  6.83  &   7.26   &  10.72 &   3.57 &  3.68  &   6.10 \\
C$^2$DSR(GRU4Rec) &  7.68  &   8.02   &  11.60 &   3.92 &  4.02  &   6.54 \\
\midrule
SASRec &   7.30 &  7.79   &   11.68 &   3.79 &  3.93 &   6.62 \\
SASRec++ &  8.09  &   8.70   &  13.11 &   4.20 &  4.45  &   7.55 \\
C$^2$DSR$^*$ &  8.73  &   9.36   &  13.66 &   4.41 &  4.55 &   7.70 \\
C$^2$DSR &  8.91  &   9.71   &  14.54 &   4.65 &  4.94 &   8.18 \\

\bottomrule
\end{tabular}
}
\end{table}

\subsection{Hyperparameter Analysis}
This section investigates the parameter sensitivity of the harmonic factor $\lambda$ and graph neural network depth $L$.

\begin{figure}[t]
	\begin{center}
		\includegraphics[width=8.5cm,height=4cm]{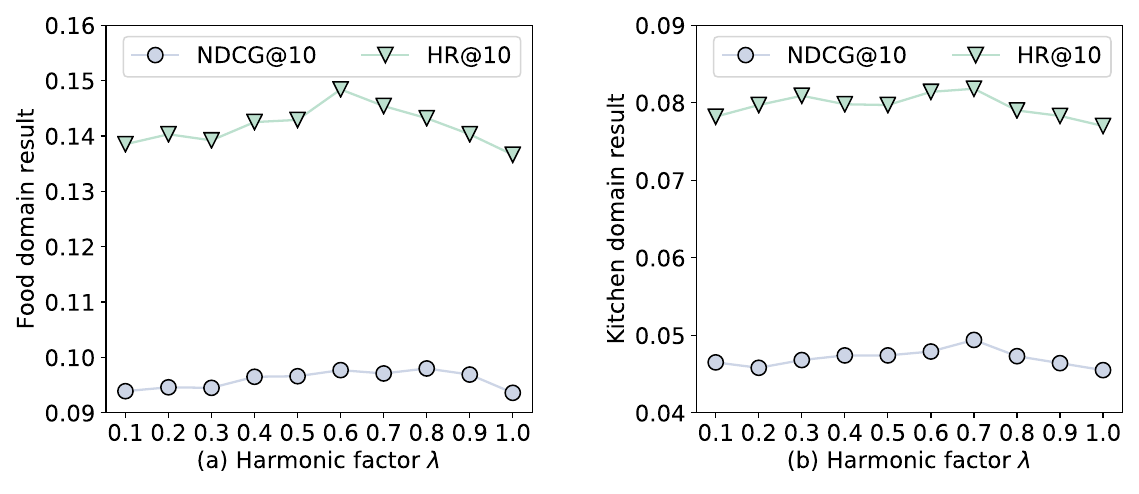}
		\caption{Result of harmonic factor $\lambda$.}
		\label{lambda_para}
	\end{center}
\end{figure}

For hyperparameter $\lambda$, the Figure~\ref{lambda_para}(a) and~\ref{lambda_para}(b) show its ``Food'' domain and ``Kitchen'' domain prediction performance in terms of NDCG@10 and HR@10, respectively.
According to it, we can find that our model achieves consistent improvements than $\lambda = 1.0$ (This case means trained without our contrastive infomax objective).
Specifically, our model shows superior and robust results when the $\lambda$  is in the range $[0.3, 0.7]$.
Even in the worst settings, our C$^2$DSR is still better than other baselines shown in Table \ref{foodkitchen}.
Besides, we find that setting $\lambda$ as a larger number could lead to faster speed for training convergence.
That is to say, $\lambda = 0.7$ might be a balanced choice between the model effectiveness and model efficiency.

\begin{figure}[t]
	\begin{center}
		\includegraphics[width=8.5cm,height=4cm]{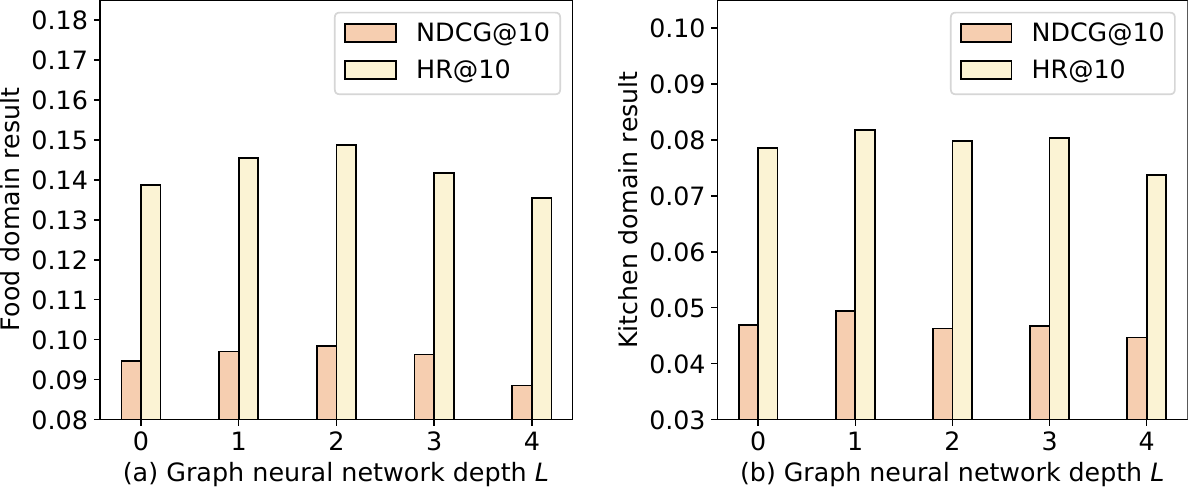}
		\caption{Impact of GNN depth $L$.}
		\label{GNN_para}
	\end{center}
\end{figure}

For hyperparameter $L$, the Figure~\ref{GNN_para}(a) and~\ref{GNN_para}(b) show its ``Food'' domain and ``Kitchen'' domain prediction performance in terms of NDCG@10 and HR@10, respectively.
As shown in this Figure, we report the recommendation results under the $L = \{0, 1, 2, 3, 4\}$, note that the $L = 0$ means ignoring the graph neural network module.
Generally, our model gives steady improvements in the cases $L = \{1,2\}$ and shows the degeneration performance in the cases $L = \{3,4\}$.
The reason might be that a deep graph neural network easily causes the over-smoothing issue, which limits the model effectiveness to capture collaborative signal. 
Therefore, choosing smaller values of $L$ is a reasonable way.

\section{Related Works}

\noindent\textbf{Contrastive Learning} provides a promising paradigm to measure the dependency of input variables by calculating their mutual information (MI).
In past years, several classical works have proposed to utilize contrastive learning for computer vision, neural language processing, and graph data, such as DIM~\cite{dim}, CPC~\cite{CPC} and DGI~\cite{dgi}.
The DIM develops a contrastive objective to maximize the MI between local and global image representation.
The CPC is designed for structure sequence data, which maximizes the MI between partial past representations and its summarized future representations.
The DGI further extended contrastive learning to unstructured graph data, which maximizes the MI between local sub-graph representations and corresponding global graph representations.
Recently, this idea to the sequential recommendation, such as S$^3$-Rec~\cite{s3rec} proposes contrastive objectives to maximize the mutual information of item context information in different forms or granularities.
In this work, we devise a novel contrastive infomax objective to maximize the MI between the single- and cross- domain representations for CDSR.

\noindent\textbf{Cross-Domain Recommendation} is a powerful technique to alleviate data sparsity issue in recommender system. 
Typical cross domain recommendation models are extended from single-domain recommendation models, such as the CMF~\cite{cmf}, and M$^3$Rec~\cite{m3rec}. 
Those methods exploit the interactions from other auxiliary domains to fulfill the user behaviours to make better recommendation in target domain.
Recently, the idea of transfer learning motivates many efforts, such as CoNet~\cite{conet}, and BiTGCF~\cite{bitg}. 
Those methods first utilize two base neural networks to model user-item interactions in source and target domain separately, and then devise distinct transfer modules to fuse them.
On top of that, latest works are focus on mining and transferring the domain-shared information across domains, such as DisenCDR~\cite{disencdr} and CDRIB~\cite{cdrib}.

\noindent\textbf{Sequential Recommendation} models the dynamic user preference to predict future items by latest historical interaction sequences. 
The pioneering works of SR are always based on the Markov Chain assumption~\cite{markovsr,transmarkov,markov}.
Those methods learn an item-item co-occurrence relationship and utilize it to predict the next item given the last interacted items. 
Recently, with the wave of neural networks, various methods have been proposed to model SR, such as recurrent networks (GRU4Rec~\cite{gru4rec}, HRNN~\cite{hrnn}), convolutional neural networks (Caser~\cite{caser}), attention networks (SASRec~\cite{sasrec}, NARM~\cite{narm}), and graph networks (SR-GNN~\cite{srgnn}).

\noindent\textbf{Cross-Domain Sequential Recommendation} aims to make better recommendation for SR task that the items belong to several domains. 
The pioneering work is the $\pi$-Net~\cite{pinet} and PSJNet~\cite{PSJnet}, those methods devise elaborate gating mechanisms to transfer the single-domain information.
\citeauthor{Zhuangict}~\cite{Zhuangict} propose a cross-domain novelty-seeking algorithm for better modeling an individual’s propensity for knowledge transfer across different domains.
Besides, some works construct interacted graph (or knowledge graph) between different domain items to guide the information transferring across domain, such as DA-GCN~\cite{dagcn} and MIFN~\cite{mifn}.
In industry, the SEMI~\cite{kddsemi} and RecGURU~\cite{recguru} are proposed  for cross-domain short-video recommendation (e.g., Taobao Video and Tencent Video), by utilizing the additional multi-modal information (e.g., text and key-frame information) or the adversarial Learning.
Compared with these methods, our C$^2$DSR has important design differences as follows: (1) we consider the single- and cross- domain user preferences at same time, (2) we devise a novel contrastive infomax objective to capture the cross-domain correlation.

\section{Conclusion}
This paper proposes a novel model C$^2$DSR for cross-domain sequential recommendation, which simultaneously leverages the intra- and inter- sequence item relationships, and jointly learns the single- and cross- domain user preferences. 
Particularly, our method includes a graphical and attentional encoder to leverage the intra- and inter- sequence item relationships simultaneously, and two sequential objectives with a contrastive objective to jointly learn and enhance the single- and cross- domain user preferences.
Additionally, we release the re-splited datasets for a fair comparison.
Empirical results demonstrate the effectiveness of C$^2$DSR, reaching a new state-of-the-art performance.
Besides, we analyze the effectiveness of our model components and the contrastive objective in detail.
In the future, we will explore our model in multi-domain setting and in continuous time space.

\section*{Acknowledgement}
We are very grateful to Jie Yu for helpful discussions on early draft.
We are also thank anonymous reviewers for their constructive comments.
This work was supported by the National Key Research and Development Program of China under Grant No.2021YFB3100600, the Strategic Priority Research Program of Chinese Academy of Sciences under Grant No.XDC02040400, and the Youth Innovation Promotion Association of CAS under Grant No.2021153.

\balance
\bibliographystyle{ACM-Reference-Format}
\bibliography{sample-base-extend.bib}
\end{document}